\documentclass[%
preprint,
superscriptaddress,
groupedaddress,
 amsmath,amssymb,
 aps,
longbibliography,
]{revtex4-1}

\usepackage{graphicx}
\usepackage{dcolumn}
\usepackage{bm}
\usepackage{multirow}
\usepackage[mathlines]{lineno}

\usepackage{siunitx}
\usepackage{ulem}
\usepackage{soul}

\begin{document}

\title{Snowmass Whitepaper AF6: Plasma-Based Particle Sources}

\author{M. Fuchs}
\email[Corresponding author: ]{mfuchs@unl.edu}
\author{B. A. Shadwick}
\affiliation{Department of Physics and Astronomy, University of Nebraska, Lincoln, Nebraska 68588, USA}
\author{N. Vafaei-Najafabadi}
\affiliation{Stony Brook University, Department of Physics and Astronomy, Stony Brook, New York 11794, USA}
\author{A. G. R. Thomas}
\affiliation{G\`erard Mourou Center for Ultrafast Optical Sciences and Department of Nuclear Engineering and Radiological Sciences, University of Michigan, Ann Arbor, MI 48109, USA}
\author{G. Andonian}
\affiliation{Department of Physics and Astronomy, University of California, Los Angeles, California 90095, USA}
\author{M. B\"uscher}
\affiliation{Peter Gr\"unberg Institut (PGI-6), Forschungszentrum J\"ulich, J\"ulich, Germany and \\
 Institut für Laser- und Plasmaphysik, Heinrich-Heine-Universität D\"usseldorf, D\"usseldorf, Germany}
\author{A. Lehrach}
\affiliation{JARA-FAME (Forces and Matter Experiments), Forschungszentrum J\"ulich and RWTH Aachen University, Aachen, Germany and \\
 Institut für Kernphysik (IKP-4), Forschungszentrum J\"ulich, J\"ulich, Germany}
\author{O. Apsimon}
\affiliation{The University of Liverpool, Liverpool L69 3BX, United Kingdom}
\author{G. Xia}
\affiliation{University of Manchester, M13 9PL, Manchester, United Kingdom}
\author{D. Filippetto, C. B. Schroeder}
\affiliation{Lawrence Berkeley National Laboratory, Berkeley, California 94720, USA}
\author{M. C. Downer}
\affiliation{Department of Physics, University of Texas, Austin, Texas 78712, USA}

\maketitle

\tableofcontents
\makeatletter
\let\toc@pre\relax
\let\toc@post\relax
\makeatother

\newpage
\section{Executive Summary}

High-brightness beams generated by particle injectors and particle sources based on advanced accelerator concepts have the potential to become an essential part of future accelerator technology. In particular, high-gradient accelerators can generate and rapidly accelerate particle beams to relativistic energies. The rapid acceleration and strong confining fields can minimize irreversible detrimental effects to the beam brightness that occur at low beam energies, such as emittance growth or pulse elongation caused by space charge forces. Due to the high accelerating gradients, these novel accelerators are also significantly more compact than conventional technology. Advanced accelerators can be extremely variable and are capable of generating particle beams with vastly different properties using the same driver and setup. So far, efforts have mainly been focused on the generation of electron beams, but there are concepts to extend the sources to generate spin-polarized electron beams or positrons. 

The beam parameters of these particle sources are largely determined by the injection and subsequent acceleration processes. Although, over the last decade there has been significant progress in the demonstrated beam parameters, oftentimes these are not in combination with other crucial parameters that are required for a future collider or more near-term applications, including X-ray free-electron lasers (XFELs), such as a sufficiently small energy spread and small emittance for bunches with a high charge and at high pulse repetition rate. 

Major research and development efforts are required to realize these approaches for a front-end injector for a future collider in order to address these limitations. In particular, this includes methods to control and manipulate the phase-space and spin degrees-of-freedom of ultrashort LWFA electron bunches with high accuracy, methods that increase the laser-to-electron beam efficiency and increased repetition rate. This also includes the development of high-resolution diagnostics, such as full 6D phase-space measurements, beam polarimetry and high-fidelity simulation tools.

A further increase in beam luminosity can be achieve through emittance damping. Emittance cooling via the emission of synchrotron radiation using current technology requires kilometer-scale damping rings. For future colliders, the damping rings might be replaced by a substantially more compact plasma-based approach. Here, plasma wigglers with significantly stronger magnetic fields are used instead of permanent-magnet based wigglers to achieve similar damping performance but over a two orders of magnitude reduced length.

\section{\label{sec:intro}Introduction}

Novel advanced accelerators have the potential to become an essential part of future accelerator technology. This includes plasma-based and advanced structure accelerators. In particular, laser-wakefield accelerators (LWFA), which are based on laser-plasma interactions, can produce high-brightness femtosecond electron bunches with low transverse emittance (for current state-of-the art parameters see Table \ref{Table:parameters}). However, some parameters of LWFA electron beams, such as the transverse emittance, energy spread, laser-to-beam conversion efficiency or pulse repetition rate can be even further improved and more specifically tailored to their application, such as colliders (this is discussed in separate white papers as part of these proceedings~\cite{LWFA-WP,SFWA-WP,PWFA-WP}) or free-electron lasers (FELs), which is discussed in a separate white paper in these proceedings~\cite{nearterm-WP}.

The parameters of LWFA electron beams are primarily determined by the electron injection into the accelerating plasma structure and the acceleration process itself. For simplicity, the majority of current LWFAs use a self-injection scheme, which leads to electron beams with a relatively large energy spread and that are typically less reproducible compared to conventional accelerator technology. To achieve electron self-injection requires a comparably high laser intensity, which is currently limiting the repetition rate at which LWFA beams can be produced. Different schemes of controlled injection have been demonstrated, including colliding laser pulses~\cite{Faure:2006a}, plasma density modulations~\cite{Geddes:2008a,Schmid:2010a} and ionization injection~\cite{McGuffey:2010a,Pak:2010a,Clayton:2010a}. First promising results have been obtained and the methods have helped to significantly improve the beam parameters over the last decade (see Table~\ref{Table:parameters}). However, these methods still require further improvements and some experimental implementations are challenging. 

The full characterization of the 6D bunch phase space distribution is extremely challenging and because of its ultrashort duration, so far it has been mainly only possible to measure bunch-integrated properties, such as energy spread and transverse emittance. However, there are measurements that indicate that some of the local bunch properties (slice emittance, slice energy spread) are smaller than that of the overall bunch and that they might be temporally correlated. For example, this includes the observation of micro-bunching of LWFA bunches at optical wavelengths \cite{Lumpkin:2020}, the demonstration of energy-chirp compensation through a tailored plasma density \cite{Doepp:2018a}, and the observation of exponential amplification of a laser-driven free-electron laser (FEL) \cite{Wang:2021a}.

Novel approaches for an injector front-end for a future plasma-based collider are required to address these limitations. In particular, this includes methods that increase the laser-to-electron beam efficiency, enable shaping the phase-space of ultrashort LWFA electron bunches with high accuracy and high-resolution diagnostics over a wide range parameter range.


\section{\label{sec:background}Background}
\subsection{Laser-Plasma Driven Sources}
\subsubsection*{LWFA in the Bubble Regime}
One of the main challenges for LWFAs are improvements in the electron beam brightness, including the energy spread and transverse emittance and increasing the accelerator repetition rate. Due to simplicity, the majority of LWFAs are driven in the highly nonlinear (bubble) regime and use electron self-injection~\cite{Pukhov:2006a,Kostyukov:2009a,Kalmykov:2009a}. Operating in this regime requires laser pulses with relativistic intensities. Specifically, the normalized vector potential $a_0=eE\lambda/2\pi mc^2 \simeq \lambda [\mu \mathrm m] (I_0 [\mathrm {W/cm}^2]/1.4 \times 10^{18} )^{1/2}$  of the pulse, where $E$ is the laser electric field, $\lambda$ the laser wavelength, $mc^2$ the electron rest mass and $I_0$ the laser intensity,  needs to be significantly in excess of 1. Furthermore, the laser pulse duration has to be significantly shorter than the plasma wavelength. To achieve these intensities with a matched laser spot size requires laser pulses with a power of hundreds of terawatts to petawatts \cite{Pukhov:2006a,Tajima:1979a,Lu:2006a}. This currently limits the repetition rate at which state-of-the art laser systems can operate, thus limiting the repetition rate of the accelerator. Note that the development of lasers with a high peak and a high average power is discussed in a separate white paper as part of these proceedings~\cite{laser-WP}. Furthermore, the highly nonlinear bubble regime increases the difficulty of control over the electron beam properties. This includes the control of the injection process and over the acceleration process. For the latter, this is because of the evolution of the laser pulse due to the laser-plasma interaction. This also leads to comparably low laser-to-electron-bunch efficiencies because of the stronger interaction of the laser with the plasma at higher intensities. 

\subsubsection*{Electron Injection in LWFAs}
Self-injected electron beams have a finite transverse emittance because the injected electrons have an intrinsic transverse momentum at the time of injection \cite{Pukhov:2006a,Kostyukov:2009a,Kalmykov:2009a}. In this case, a laser pulse with a sufficiently high laser intensity ponderomotively excites a plasma wave by transversally expelling plasma background electrons, leaving behind a fully evacuated ionic cavity (the bubble). The expelled electrons are attracted back towards the axis by the electrostatic fields due to the ion cavity. Most electrons that are transversely expelled by the laser from a region close to the axis wrap around the cavity in half circles. They compose a highly dense electron sheath around the bubble center. Electrons within a specific initial off-axis region propagate along trajectories where they obtain a sufficiently large longitudinal momentum to become trapped in the bubble. The bubble structure can become unstable due to extensive beam loading or because of laser pulse alterations through self-evolution of the pulse in the plasma. This can lead to an extension of the bubble, leading to subsequent electron injection and ultimately a decrease in beam quality~\cite{Kalmykov:2009a}.

Parameters of the electron beams can be controlled via injection mechanisms. The currently mainly used processes include (i) colliding laser pulse injection, (ii) ionization injection and (iii) shockfront-assisted injection. In the colliding laser pulse injection, two laser pulses with the same polarization collide, each with an intensity below the self-injection threshold \cite{Faure:2006a}. The colliding pulses generate a beatwave, which allows background plasma electrons to cross the separatrix and become trapped. The manipulation of the electron energy and a reduction in the electron energy spread to approximately 1\% have experimentally been demonstrated. Ionization injection uses a gas mixture of low-Z and higher-Z atoms. As the ionization of the inner-shell electrons of the higher-Z atoms occurs during higher intensity part of the laser pulse, they can be born and injected into a suitable acceleration phase of the bubble \cite{McGuffey:2010a,Pak:2010a,Clayton:2010a}. This can lead to a decrease in transverse emittance and a localized injection along the accelerator. A variation on the ionization injection scheme uses two lasers with a large difference in wavelength. Here, the long-wavelength laser (e.g. a CO$_2$ laser at $\lambda =10~ \mu$m) drives the wakefield and the short-wavelength laser (e.g. a Ti:Sapph laser at $\lambda \sim 1~ \mu$m) ionizes and injects the electrons into the wake. Because of the scaling of the normalized vector potential $a_0^2 \propto I\lambda^2$, the long-wavelength laser can achieve a high vector potential at a relatively low intensity as compared to a shorter-wavelength laser. As a result, the long-wavelength driver does not fully ionize the gas target and the short-wavelength injector pulse can ionize and specifically inject the remaining inner-shell electrons. LWFAs driven by long-wavelengths lasers are typically driven using a lower plasma densities and in plasma bubbles with a significantly bigger volume. This relaxes the required precision for injection and can help maintaining spin polarization and low energy spread.This "two-color ionization injection" scheme has been shown in simulations to produce beams with an emittance that is low enough to meet the requirements of a collider \cite{Xu:2014,Yu:2014}.
In shockfront-assisted injection, electron injection is controlled via a longitudinal plasma density downramp inside the gas target \cite{Geddes:2004a,Buck:2013a}. As the plasma wave propagates through a density downramp, its local phase velocity decreases. It can be reduced to approximately the plasma fluid velocity which leads to the injection of cold background plasma electrons.

\section{Key results since last Snowmass and their relevance to high energy physics goals}
\subsection{Laser-Plasma Driven Sources}
The parameters of LWFA beams have tremendously improved over the last decade in terms of beam energy, accelerated charge, energy spread and repetition rate (see Table \ref{Table:parameters}). Unlike previous decades when the beam improvements have heavily relied on advancements in laser technology, many of these improvements are due to improved injector and accelerator designs. The ultimate limits of these technologies still needs to be explored.

\begin{table}[]
\begin{tabular}{|l|l|l|l|}
\hline
\textbf{Bunch property}                                                                                   & \textbf{State of the Art}                                              & \textbf{Other beam parameters}                                                              & \textbf{References}                                                           \\ \hline \hline
\textbf{Bunch energy}                                                                                     & 8 GeV                                                                  & \begin{tabular}[c]{@{}l@{}}5 pC, 0.2 mrad\\ (up to 60 pC in 6 GeV peak)\end{tabular}        & \begin{tabular}[c]{@{}l@{}}Gonsalves \textit{et al.}, \\ PRL (2019)\cite{Gonsalves_2019}\end{tabular}       \\ \hline
\multirow{3}{*}{\textbf{\begin{tabular}[c]{@{}l@{}}Bunch charge\\ \\ \end{tabular}}} & \begin{tabular}[c]{@{}l@{}}220 pC\\ (dE/E = 14\% FWHM*)\end{tabular}    & \begin{tabular}[c]{@{}l@{}}250 MeV, 7 mrad\\ {[}ionization injection{]}\end{tabular}        & \begin{tabular}[c]{@{}l@{}}Couperus \textit{et al.}, \\ Nat. Comm. (2017)~\cite{Couperus_2017}\end{tabular} \\ \cline{2-4} 
                                                                                                          & \begin{tabular}[c]{@{}l@{}}338 pC\\ (dE/E = 15\% FWHM*)\end{tabular}    & \begin{tabular}[c]{@{}l@{}}216 MeV, 0.36 mrad\\ {[}shock front injection{]}\end{tabular}    & \begin{tabular}[c]{@{}l@{}}Götzfried \textit{et al.}, \\ PRX (2020)~\cite{Goetzfried_2020}\end{tabular}       \\ \cline{2-4} 
                                                                                                          & \begin{tabular}[c]{@{}l@{}}700 nC\\ (dE/E = 100\%*)\end{tabular}        & \begin{tabular}[c]{@{}l@{}}Up to 200 MeV\\ laser: OMEGA-EP, 100 J, 700 fs\end{tabular}      & \begin{tabular}[c]{@{}l@{}}Shaw, \textit{et al.} \\ Sci Rep 11 (2021)~\cite{Shaw:2021}\end{tabular}     \\ \hline
\textbf{Energy spread*}                                                                                    & 0.2 – 0.4\% (RMS)                                                      & \begin{tabular}[c]{@{}l@{}}800 MeV, 8.5 – 24 pC\\ shockwave assisted injection\end{tabular} & \begin{tabular}[c]{@{}l@{}}Ke, \textit{et al.} \\ PRL (2021)~\cite{Ke:2021a}\end{tabular}              \\ \hline
\multirow{2}{*}{\textbf{Bunch duration}}                                                                  & 1.4 fs (RMS)                                                           & \begin{tabular}[c]{@{}l@{}}15 pC, CTR \\ (diagnostic limited)\end{tabular}                  & \begin{tabular}[c]{@{}l@{}}Lundh \textit{et al.}, \\ Nat Phys (2011)~\cite{Lundh:2011a}\end{tabular}       \\ \cline{2-4} 
                                                                                                          & 2.5 fs (RMS)                                                           & \begin{tabular}[c]{@{}l@{}}Faraday rotation \\ (diagnostic limited)\end{tabular}            & \begin{tabular}[c]{@{}l@{}}Buck \textit{et al.}, \\ Nat Phys (2011)~\cite{Buck:2011a}\end{tabular}       \\ \hline
\textbf{\begin{tabular}[c]{@{}l@{}}Emittance*\\ (normalized)\end{tabular}}                                 & \begin{tabular}[c]{@{}l@{}}0.2 $\pi$ mm mrad\\ (@245 MeV)\end{tabular} & Single-shot measurement                                                                     & \begin{tabular}[c]{@{}l@{}}Weingartner \textit{et al.}\\ PRSTAB (2012)~\cite{Weingartner:2012a}\end{tabular}    \\ \hline
\multirow{2}{*}{\textbf{Repetition Rate}}                                                                 & 1 Hz                                                                   & \begin{tabular}[c]{@{}l@{}}24-hour operation; \\ 100,000 consecutive shots\end{tabular}     & \begin{tabular}[c]{@{}l@{}}Maier \textit{et al.}, \\ PRX (2020)\cite{Maier_2020}\end{tabular}            \\ \cline{2-4} 
                                                                                                          & 1 kHz                                                                  & up to 15 MeV, 2.5 pC                                                                        & \begin{tabular}[c]{@{}l@{}}Salehi \textit{et al.}, \\ PRX (2021)~\cite{Salehi_2021}\end{tabular}           \\ \hline
\textbf{\begin{tabular}[c]{@{}l@{}}Efficiency\\ (laser-to-electron)\end{tabular}}                         & 3\%                                                                    & 2J in driver laser pulse                                                                    & \begin{tabular}[c]{@{}l@{}}Götzfried \textit{et al.}, \\ PRX (2020)~\cite{Goetzfried_2020}\end{tabular}       \\ \hline \hline
\end{tabular}
\caption{\textbf{Overview of the state-of-the art LWFA electron beam parameters.} \\ *bunch-integrated measurements}
\label{Table:parameters}
\end{table}

\section{\label{sec:concepts}Proposed concepts and development path}
\subsection{Laser-Plasma Driven Sources}
While LWFAs have demonstrated beams with high brightness, many of the parameters listed in Table \ref{Table:parameters} have not been realized simultaneously. In addition to further improvements in each of these parameters, methods to control the beam phase-space that allow the generation of high-brightness beams that combine multiple of these record bunch properties are needed. This not only requires the development of new injection and control methods but also the development of diagnostics with high spatial and temporal resolution that allow measurement of the bunch parameters with sub-bunch length precision. Furthermore, it requires a combined experimental and theory efforts that includes the development of novel experimental and simulation methods. High-resolution diagnostics will allow the comparison of experimental results with simulation to a high degree and can help improve the ability for predictions based on simulations. 

Finally, the requirements on the repetition rate of the driver laser to achieve a sufficiently high beam luminosity also requires the determination of the properties for a best-suited driver laser, which are likely going to be different for an injector front-end and the subsequent acceleration stages. This also includes developing methods to increase the efficiency.

\subsubsection*{Beam Control and Phase-Space Shaping}
The generation of LWFA electron bunches with a sufficiently high beam quality requires control over the beam phase space with very high temporal and spatial precision. This includes control over both, the electron injection process and the acceleration process. This can be achieved by achieving separate control over injection and acceleration processes by decoupling the injection from the acceleration stage. As described above, this has been demonstrated and has resulted in improvements in the LWFA parameters. 

Efforts to shape the electron bunch phase-space include the development of advanced methods of electron injection including beam tapering and the investigation of acceleration in different regimes, such as a more linear regime \cite{Schroeder:2010}. This includes further injection schemes that have the potential for more control, such as using two, multi-color laser pulses \cite{Yu:2014} or laser pulses with higher-order spatial modes \cite{Yu:2015a} that have been proposed but are lacking experimental investigation.

Also of interest are methods to control phase space including the design of sophisticated driver lasers, such as high control over the laser spatial and temporal higher-order shapes, multiple pulses, potentially of different colors or the incoherent addition of multiple pulses for example from fiber lasers and control over the laser evolution during the laser-plasma interaction.

Furthermore, novel advanced target designs have the potential to increase the beam quality and shot-to shot stability.

\subsubsection*{Increase in Efficiency and Stability}
A high laser-to-electron bunch efficiency is crucial to ensure operation at a high repetition rate. A high reproducibility of the electron bunch parameters are crucial for stable operation. The efficiency and stability of LWFAs can be improved through control over the injection process, the acceleration and the type of acceleration process. While most of the current LWFAs are driven in the highly-nonlinear bubble regime, the quasilinear regime, in which the plasma waves are driven only moderately relativistically and the wakefield is approximately sinusoidal, can lead to a more efficient acceleration \cite{Schroeder:2010}. The quasilinear regime also has the advantage that the accelerating and focusing phase regions for electrons and positrons are nearly symmetric. Unlike the bubble regime where only a single wakefield bucket is excited, multiple buckets can be driven in the quasilinear regime. This allows, for example, the acceleration of bunch trains with the advantage of optimal beam loading of each bucket. The bunch structure of the pulse train is also advantages for a future collider as the short duration of each bunch decreases Beamstrahlung effects \cite{Schroeder:2010}, while delivering an overall macrobunch with a high charge. However, this also requires advanced injection techniques to control the shape of the injected particle bunches. This scheme can be combined with using multiple driver laser pulses (multi-pulse laser wakefield acceleration) for efficient wake generation and high repetition rate operation \cite{Hooker:2014}. As described above, novel advanced target designs have the potential to increase the efficiency and the shot-to shot stability.

The subsequent acceleration process can be strongly impacted by the evolution of the driver laser pulse through laser-plasma interactions \cite{Kalmykov:2009a}. To increase the stability and efficiency requires control over the laser evolution, for example through pre-shaped laser pulses, the use of multiple laser pulses \cite{Kalmykov:2013a} and specific tailoring of the plasma density profile.

\subsubsection*{Optimal Driver Laser Properties}
The performance and properties of LWFAs is greatly impacted by the laser pulse properties. Optimized operation of different accelerator regimes, such as the bubble regime or the quasilinear regime, require specific laser pulse properties. Furthermore, as described above, advanced controlled injection and acceleration schemes require specific laser pulses, such as higher-order spatial and temporal modes, multiple laser pulses, multi-colors, incoherent addition. The requirements on the repetition rate of the driver laser to achieve a sufficiently high beam luminosity also requires the determination of a (set) of properties for a best-suited driver laser, which are likely going to be different for an injector front-end and the subsequent acceleration stages.

\subsubsection*{Diagnostics}
Despite ever-more sophisticated attempts to measure the LWFA electron bunch phase space \cite{Downer:2018}, it is still far from being fully characterized. In particular, the sub-femtosecond time resolution that is require is very challenging. The determination of the success of certain approaches requires novel diagnostic methods for both, the accelerated electron bunch and the accelerating plasma structure itself. These methods need to have a very high temporal and spatial resolution and ideally work in a single shot. Furthermore, they need to be capable of measuring the full 6D phase-space distribution of the electron bunches, including their temporal energy distribution with sub-bunch resolution. This includes high-resolution, (ideally) single-shot diagnostics of plasma and accelerating structure. 

These diagnostics will also allow a comparison of experimental results to high-fidelity simulations to validate and improve the simulation codes.

\subsubsection*{Combined experimental and theoretical/simulation efforts} 

The successful design and fielding of a plasma-based particle source will require significant advances in the current state of the art in the numerical simulation of such systems.  Currently, high fidelity numerical modeling of these systems is beyond the capabilities of even leadership-class computing facilities.  Hybrid models, allowing for the optimization of computational cost versus physical accuracy will be necessary.  That is, it is not viable to have a single global physics model in a simulation.  Each region of phase space will have to be optimally treated based on its intrinsic importance to the overall system.  In this way, computational resources can be allocation to provide uniform physical fidelity.  For example, the bubble sheath could be treated with a high-order kinetic model (either Lagrangian or Eulerian) whereas areas of the plasma further removed for the ``action'' might be treated, with sufficient accuracy, as a fluid.
Sampling ``noise'' associated with macro-particle models \cite{Evstatiev:2021vv} will have to be kept under tight control and new, low-noise algorithms may have to be developed.  Machine learning will be an essential element in this optimization process and we expect that the very optimization will, by identifying the critical regions of phase space,  provide additional insight into the details of the underlying physical processes.  To achieve this level of fidelity will require tight integration of experimental and theoretical efforts, each guiding advances in the other.  Diagnostics sensitive to details of phase space will be required to probe simulations results with enough precision to guide model development and the computations cost optimization process.  Macroscopic diagnostics such as energy spread and emittance are not likely sufficient constraints on simulation models.  In addition, precise characterization of the gas jet profile and incident laser pulse will be necessary to provide sufficiently accurate simulations \cite{Ferri:2016ab}.  As a consequence of the model optimization process, we expect to derived to be high-performance reduced models that allow for predictive simulations to run concurrently with experimental campaigns without requiring leadership-class computing systems.  Such a capability could provide the tight-coupling feedback needed to quickly advance the experimental program.


\subsubsection*{Polarized Electrons}

The acceleration of polarized electron beams by means of laser-driven acceleration promises to be cost-efficient and highly effective. Before a technical implementation can be envisaged, some principal issues need to be addressed theoretically, for example: (i) is it possible to alter the polarization of an initially unpolarized target through interaction with relativistic laser pulses or (ii) are the spins so inert during the short acceleration period that a pre-polarized target is required (see Ref.~\cite{Buescher_2020a} for a recent review)? Starting from the work by Hützen \textit{et al.} \cite{Huetzen_2019a} (which proposes the use of pre-polarized targets for proton acceleration), Wu \textit{et al.} \cite{Wu_2019a} and Wen \textit{et al.} \cite{Wen2019a} have developed a scheme to generate intense polarized electron beams via the interaction of an accelerating laser pulse with a pre-polarized plasma, which is produced through photo-dissociation of a dense Halide (e.g. HCl) gas-jet by a circularly polarized ultra-violet (UV) laser pulse. A specifically configured ionization injection scheme has also been shown in simulations to produce significant populations of polarized electrons \cite{Nie:2021}. Obviously, for positrons (which are not discussed here in detail) other approaches are required, see e.g. Ref.~\cite{Buescher_2020a}. These mostly rely on spin-selective radiation reactions of pre-accelerated, unpolarized electron beams with ultra-intense laser pulses. 

\begin{figure}
    \centering
    \includegraphics[width=\textwidth]{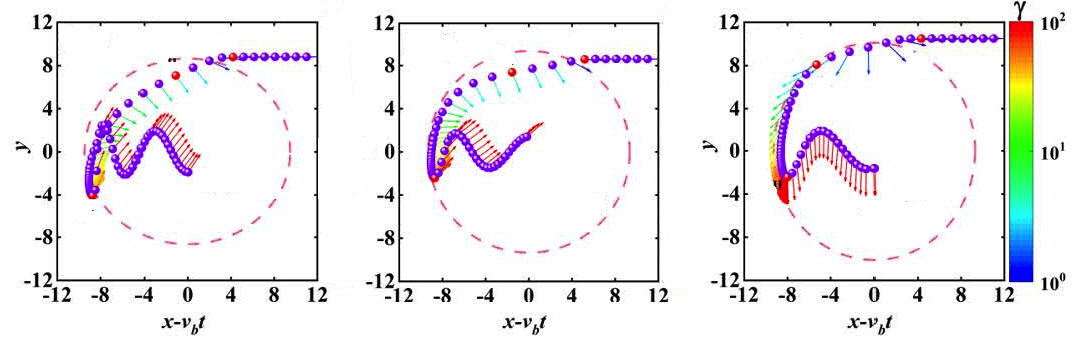}
    \caption{Temporal development of the spin orientations (arrows with color-coded electron energies) of test electrons (pink dots) during bubble acceleration (edited Fig.1 from Ref. \cite{Fan2022a}). It can be seen that slight changes of the bubble geometry (prolate, spherical, oblate; indicated by the dashed red lines) leads to  large differences of the originally longitudinally polarized electrons. The accelerating laser pulse propagates the simulation box from left to right (not shown in the figure).}
    \label{fig:polarization-1}
\end{figure}

Detailed theoretical studies (like the ones in Ref.~\cite{Fan2022a}) reveal that intense highly-polarized electron beams can be accelerated to multi-MeV energies via the ``standard" bubble mechanism with 100-TW class lasers. It has been shown that the final spin direction strongly depends on the self-injection process (see Fig.\ref{fig:polarization-1}) and, thus, a careful tuning and control of the laser/target parameters is mandatory. As a consequence plasma-based accelerators promise to easily adapt the degree of polarization and the preferred spin direction according to the experimental needs (e.g. longitudinal or transversal polarization). Once  energies above a few MeV have been reached, the beam polarizations are very robust during  post-acceleration in subsequent plasma stages \cite{Thomas2020a}. 

On the experimental side, the polarization of protons accelerated from an unpolarized foil target has been measured at the Arcturus laser at Düsseldorf \cite{Raab2014a}.
A polarized $^3$He gas-jet target \cite{Fedorets2022a} that has been used for a first experimental campaign at the Phelix laser facility at GSI Darmstadt. 
A polarized HCl target for proton acceleration has been prepared at Forschungszentrum J\"ulich \cite{Huetzen_2019a}. It is planned to upgrade this target to deliver also polarized electrons. 
On the European scale, similar studies are under way in the framework of the EuPRAXIA consortium
\cite{Eupraxia2020a}.

\subsubsection*{Advanced Transverse Emittance Cooling Technique}

Over the last few decades, two different conventional design approaches, using normal conducting (CLIC, \cite{CLIC}) and superconducting (ILC, \cite{ILC}) radio frequency structures, revealed the size of such a machine to be about 30$-$50$\,$km. The electron-electron option of the Future Circular Collider (FCC) project envisages a 100\,km circumference machine for collisions at centre-of-mass energy of 90\,GeV$-$350\,TeV \cite{FCC}. The limits of conventional technologies for high energy demand of particle physics exploration persuade us to new frontiers of particle accelerator science. Paradigm shifting technologies are being developed such as plasma acceleration \cite{8GeV, awake, energy_doubling}. However, there are still many challenges to be addressed before the maturation of plasma technology for large-scale accelerator applications. Advanced and novel accelerators community (ICFA- ANAR2017) underlined and prioritised the following technological challenges; the repetition rate, efficiency and beam quality of the lasers; scalability of the system; delivery of high collision luminosity; resolution of diagnostics; comprehensive simulations and the availability of dedicated test facilities \cite{ANAR19}. 

In order to achieve high luminosity, the ILC design relies on emittance cooling via damping rings followed by a long transfer line from ring to main linear accelerator followed by beam delivery system to the final interaction point. Emittance cooling is achieved through radiation damping due to synchrotron radiation emitted from beam particles moving along curved trajectories of a circular accelerator (damping ring) that has a circumference of a few hundred meters to kilometers. Particles emit synchrotron radiation depending on the local curvature of the orbit within a cone of angle $\/\gamma$, where $\gamma$ is the relativistic Lorentz factor. Longitudinal momentum is restored by radio-frequency cavities in the ring, while transverse momentum is damped on every turn through radiation damping and quantum excitation are equal and an equilibrium value is reached \cite{handbook}, typically in milliseconds \cite{ILC_damping_time, sys_damp_design}. A faster damping is achieved by increasing the energy loss per turn by adding high field periodic magnetic structures (wigglers or wigglers, depending on the magnetic strength) \cite{wiggler_damping}. 

As an example, ILC project proposes a machine with 250-500$\,$GeV centre-of-mass energy over about a 31$\,$km footprint \cite{ILC_tech_des_report}. Electrons and positrons emerging from different sources undergo an initial acceleration up to 5$\,$GeV before they are injected into a their respective damping rings with a circumference of 3.2$\,$km, housed in the same tunnel. ILC damping rings are designed in a race track shape to accommodate two straight sections. A radiative section comprising 54 super-ferric wigglers is located in one of these straight sections. Each wiggler is 2.1$\,$m long and generates a 2.16$\,$T peak magnetic field when operating at 4.5$\,$K and radiates 17 kW radiation power \cite{ILC_tech_des_report}. This straight section also houses a superconducting radio-frequency system to replenish the longitudinal momentum of the beam. 

An innovative cooling method was envisaged which shows great promise of replacing the magnetic wigglers in a damping ring with plasma wigglers and providing superior beam quality for the future linear collider based on plasma technology. A compact synchrotron radiation insertion device with significantly larger fields than the current wigglers would limit the footprint and the cost of the ring by reducing the required number and size of these insertion devices. There are several methods for plasma assisted radiation generation from a relativistic particle beam, such as, betatron oscillations, Compton scattering, bremsstrahlung and transition radiation \cite{Alec2016}. We propose to incorporate the concept of plasma wiggler as radiators in a damping ring to benefit from their large effective magnetic fields and compactness. There are various concepts to conceive a plasma wiggler \cite{PRL114,plasma_undulator1,plasma_undulator2,plasma_undulator3}. Following the one proposed in \cite{PRL114}  , a plasma wiggler is formed when a short laser pulse is injected into plasma off-axis or at an angle that causes the centroid of the laser pulse to oscillate. Given that the product of the plasma wave number and the characteristic Rayleigh length of the laser is much larger than one, the ponderomotively driven plasma wake will follow this centroid. This oscillating transverse wakefield works as an wiggler forcing particles to follow sinusoidal trajectories and emit synchrotron radiation. In addition, the damping time is inversely proportional to the square of the magnetic field of the damping device. It is numerically demonstrated that a plasma wiggler can generate order of magnitude larger effective magnetic fields than conventional wigglers, hence can reduce the length of the damping units by a factor of hundred while providing the same damping times \cite{Icarus_arxiv}.

\subsubsection*{Plasma Photocathode in Beam-Driven Wakefield}
In the plasma photocathode concept,  electrons are generated directly within a beam-driven plasma wakefield accelerator, by the ionization of neutral gas particles using a high intensity laser pulse \cite{Hidding:2012prl}.
The electric fields generated in a plasma wakefield accelerator can exceed those of traditional photoinjectors by many orders of magnitude.
Beam emittance is dependent on the injection properties of the plasma electrons into the accelerating phase of the plasma wave.
Once injected at the proper phase of the plasma wave, the electrons are subject to $\sim$GV/m accelerating fields, as well as focusing electric fields inside the plasma blowout, reducing the space charge effects that typically lead to an increase in the beam emittance.
The plasma photocathode thus offers a path for the generation of very high brightness beams.

The decoupling of beam injection and acceleration is accomplished using different plasma sections, such as a dual species gaseous media.
In a mixture of both low-ionization-threshold and higher-ionization-threshold gas components, the higher-ionization-threshold components are still present in unionized form within a plasma cavity.
The plasma photocathode concept is based on the  release of electrons via ionization of these higher-ionization-threshold states with a focused laser pulse at an appropriate position directly within the accelerating plasma blowout.

The optically initiated injection and acceleration of electron bunches, generated in a multi-component hydrogen and helium plasma employing a spatially aligned and synchronized laser pulse, was demonstrated in an experiment at SLAC FACET \cite{Hidding:2017ipac}.
In the experiment, a pre-ionized plasma channel is formed in a hydrogen–helium gas mixture.
A 20.35GeV electron beam, with charge of 3.2nC and pulse length of 30$\mu$m, drives a  wakefield in the hydrogen plasma, but does not ionize the helium gas.
A laser pulse (800nm wavelength, 10$^{15}$W/cm$^2$ intensity) is focused within this plasma wakefield, to liberate the helium electrons.
The properties of the beam depend on the specific injection mechanism. 
When the laser pulse arrives before the drive bunch, known as the plasma torch mode, the plasma wave is distorted due to the presence of ionized helium particles.
If the laser pulse arrives directly after the drive bunch, the plasma photocathode regime is achieved, without distortion of the plasma wave \cite{Deng:2019nat}.
The two injection modes depend on the synchronization between ionization laser and drive bunch. 
The experimental demonstration of the plasma photocathode is a significant milestone and offers a path towards the production of electron beams with nanometer-radian normalized emittances \cite{Deng:2019nat}.

The plasma photocathode concept based on plasma waves can further be modified, to relax both beam and laser requirements by using a dielectric wakefield accelerator in place of the plasma wakefield.
In this conceptual scheme, a drive beam propagates axially through the center of a dielectric lined waveguide that is filled with a neutral gas.
The beam generates a wakefield due to the retarding nature of the dielectric medium, but the beam fields do not ionize the gas. 
An incoming, co-propagating laser is focused behind the drive beam, which locally ionizes the gas, generating a witness beam similar to the plasma wave scenario.
Although the gradients provided by the dielectric wakefields are not as intense as in the plasma case, dielectrics still enable gradients on the order of GeV/m \cite{OShea:2016natp}, before the onset of breakdown, or other high-field effects \cite{OShea:2019prl, Thompson:2008prl}. 
Additionally, the fundamental accelerating mode supported in the dielectric structure is longer than the plasma wavelengths used in the previous plasma photocathode experiments, which relaxes the stringent requirements of both beam and laser properties, and synchronization required for precision injection. 
Dielectric wakefield accelerators are solid-state structures, so there is also  reduced complexity of generating, operating, containing, and characterizing a preionized plasma column and associated complexities therein. 
Experimental efforts to demonstrate the proof-of-concept are currently being undertaken at the Argonne Wakefield Accelerator \cite{Andonian:2021ipac}

\subsubsection*{Plasmonic Photocathodes}

Nano-structuring the surface of plasmonic material has been demonstrated to resonantly couple light to the vacuum-metal interface, inducing strong local near-field enhancement ~\cite{economou_surface_1969}. Upon laser illumination, a properly engineered surface support electromagnetic traveling waves confined at the metal-dielectric interface, called surface plasmon polaritons (SPP). SPP modes are driven by electron charge-density oscillations in the material, which exhibit shorter wavelengths with respect to the illuminating laser. Therefore, mediated by SPP, the optical field energy can be transported and concentrated in areas of sub-wavelength size, leading to large local field enhancement. The nanostructured surface can be precisely engineered to obtain the amplitude and phase profile required for optimized electron generation and acceleration of electron pulses. 
Although metals sustain higher losses with respect to dielectric systems, the large field enhancement obtained relaxes the requirements on the incident laser intensity for the same accelerating gradient. Furthermore metals offer strong control of the electromagnetic field at the subwavelength scale due essentially to the high index of refraction available, and are not affected by common problems beam charging. Realization of SPP nano-cavities for electron generation has been experimentally demonstrated to generate very large field enhancements~\cite{polyakov_plasmonic_2011,li_surface-plasmon_2013}. Here the nanostructure is acting as a high-Q Fabry-Perot resonator for the SPP waves, matching the speed of the (slow) surface plasmon along the surface of the structure with the incident laser wavelength. While these structures can achieve very high field enhancements, the very high power density stored locally can quickly generate damage, especially at high repetition rates. 
An alternative path is to make use of non-resonant structures. The intrinsically low quality-factor Q of the geometry results in lower field enhancement, but also decreases the energy density stored locally and, therefore, the potential for structure damage. In this configuration SPP waves travel along the surface until they are either absorbed by the surface through electron scattering or are radiated into the vacuum through surface defects. Interference between traveling SPP can also be exploited to generate large field enhancements in specific areas of the structure, not necessarily spatially coincident with a nanoscale feature. Recently, non-resonant nano-structuring of plasmonic materials has demonstrated the ability to focus light into a nanoscopic areas~\cite{durham_plasmonic_2019}. Large local enhancement of electric field can be achieved through SPP interference, leading to broadband (i.e. ultrafast response), highly confined multiphoton photoemission from a flat surface, and therefore avoiding aberrations from  curved surfaces and burning issues of tip-like photocathodes in high field environments.

\begin{figure}
\begin{center}
 \includegraphics[height=.18\textheight]{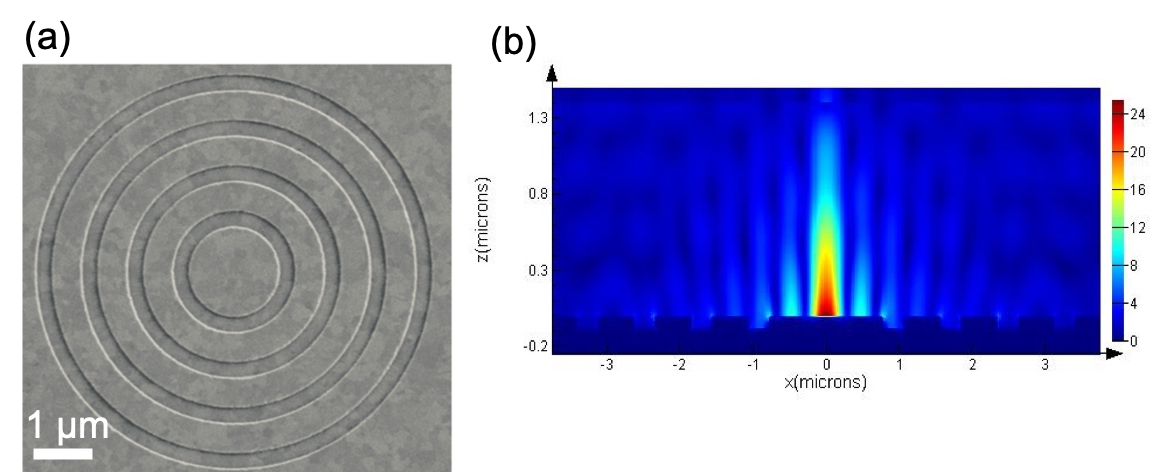}
\caption{SPP-based photocathode; (a) Structure fabricated with FIB technique; (b) distribution of field enhancement along a cut-out of the structure. From~\cite{durham_plasmonic_2019}.}
\label{applications}
\end{center}
 \end{figure}



\section{\label{sec:conclusion}Conclusion}

The injector is a critical element of a high-energy collider.  HEP-relevant Laser-plasma accelerator scenarios, require the injector to produce electron bunches with carefully tailored characteristics that have yet to be realized in the laboratory.  Developing the necessary injector technology will require a combined effort to advance the forefront of phase space diagnostics, laser technology, computational modeling of injection, and basic plasma physics.

\bibliography{references_injector}

\end{document}